\documentstyle[12pt,epsf]{article}

\def\arnps#1#2#3{  {\it Ann. Rev. Nucl. Part. Sci. }{\bf #1} (19#2) #3}
\def\npb#1#2#3{    {\it Nucl. Phys. }{\bf B #1} (19#2) #3}
\def\plb#1#2#3{    {\it Phys. Lett. }{\bf B #1} (19#2) #3}
\def\prd#1#2#3{    {\it Phys. Rev. }{\bf D #1} (19#2) #3}
\def\prep#1#2#3{   {\it Phys. Rep. }{\bf #1} (19#2) #3}
\def\prl#1#2#3{    {\it Phys. Rev. Lett. }{\bf #1} (19#2) #3}
\def\ptp#1#2#3{    {\it Prog. Theor. Phys. }{\bf #1} (19#2) #3}

\def\Im{\mathop{\mbox{Im}}}
\def\Re{\mathop{\mbox{Re}}}

\def\ss{\scriptsize}

\newcommand{\beqn}{\begin{eqnarray}}
\newcommand{\beq}{\begin{equation}}
\newcommand{\eeqn}{\end{eqnarray}}
\newcommand{\eeq}{\end{equation}}

\begin{document}
\begin{titlepage}
\begin{flushright}
{
TUM-HEP-318/98\\
July 1998
}
\end{flushright}
\vskip 0.6cm
\centerline {\Large{\bf Pre-LHC SUSY Searches: an Overview$^{*}$}}
 
\vskip 0.6cm
\centerline {A. Masiero}
\centerline {\it SISSA, Via Beirut 2-4, 34013 Trieste, Italy and}
\centerline {\it INFN, sez. di Trieste, Padriciano 99, 34012 Trieste,Italy}
\vskip 0.2cm
\centerline{\it and}
\vskip 0.2cm
\centerline {L. Silvestrini}
\centerline {\it Technische Universit\"{a}t M\"{u}nchen, Physik Department,}
\centerline {\it D-85748 Garching, Germany}

\vskip .65cm
 
\begin{abstract}
We discuss the prospects for searches of low-energy supersymmetry in the
time interval separating us from the advent of LHC. In this period of time
``indirect'' searches may play a very relevant role. We refer to
manifestations of supersymmetry in flavour changing neutral current and CP
violating phenomena and to signals of the lightest supersymmetric particle
in searches of dark matter. In the first part of the talk we critically
review the status of the minimal supersymmetric model to discuss the
chances that direct and indirect supersymmetric searches may have before
the LHC start. In the second part we point out what we consider to be the
most promising grounds where departures from the standard model prediction
may signal  the presence of new physics, possibly of supersymmetric nature.
We argue that the often invoked complementarity of direct and indirect
searches of low-energy supersymmetry is becoming even more true in the
pre-LHC era. 
\end{abstract}
\vskip 0.65cm 
\centerline{\small{$^{*}$ Invited talks given by A.
    Masiero at the Tropical Workshop on Particle}} 
\centerline{\small{Physics and Cosmology, and at the Second Latin American
    Symposium}} 
\centerline{\small{ on High Energy Physics, Puerto
    Rico, April 1-10, 1998 }}
\vfill

\end{titlepage}

\newpage
\section{Introduction}
\label{sec:intro}

It is not rare to hear the following gloomy forecast: if no
supersymmetric signal is seen at LEP, then we have nothing else to do
but wait for LHC. We do not agree with this statement. Apart from the
fact that even for direct searches one should take into account the
relevant potentialities of Tevatron in the LEP-LHC time interval, one
should not neglect that indirect searches for new physics signals are
going to be flourishing before 2005.  We refer to processes exploring
flavour physics (with or without CP violation) where new particles can
play an active role being exchanged in the loop contributions and to
several new astroparticle observations which may constitute
privileged places to obtain information on physics beyond the
Standard Model (SM).

We wish to present here a brief overview (which is necessarily biased
by our theoretical prejudices) of what we consider most promising in
this effort of looking for indirect signals of low-energy
Supersymmetry (SUSY) before the LHC advent. First we will review the
status and prospects for direct SUSY searches, then we will discuss
the role that SUSY may play in Flavour Changing Neutral Current (FCNC)
and CP violating phenomena. Finally we will briefly comment on
searches for the lightest SUSY particle in experiments looking for
Dark Matter (DM).

\section{Status of the MSSM}
\label{sec:MSSM}

It is known that, even asking for the minimal content of superfields
which are necessary to supersymmetrize the SM and imposing R parity,
one is still left with more than 100 free parameters most of which are
in the flavour sector. It is also true that very large portions on
this enormous SUSY parameter space are already ruled out by the
present phenomenology (in particular FCNC and CP constraints). If one
wants to reduce the number of free parameters one has to make
assumptions on what lies well beyond low-energy SUSY, in particular on
the quite unknown issue of the origin of SUSY breaking. The two most
popular drastic reductions of free SUSY parameters are provided by
minimal supergravity (SUGRA) \cite{SUSY1} (with the further assumption
of unification of gauge couplings and gaugino masses at some grand
unification scale) and by the models of Gauge-Mediated SUSY Breaking
(GMSB) \cite{GMSB1}-\cite{GMSB3}. In minimal SUGRA and the minimal
version of GMSB we have only 3 or 4 parameters in addition to those of
the SM and so we can become much more predictive.

In the context of the minimal supergravity model (with electroweak
radiative breaking), we ask the following relevant questions for
direct SUSY searches: i) given the present experimental lower bounds
on the masses of SUSY particles, how much room have we got in the SUSY
parameter space to explore, or, in other words, when should we give up
with SUSY if searches are fruitless? ii) is there any experimental
signature of low-energy SUSY which is independent from the choice of
the SUSY parameters in particular of the soft breaking sector? iii)
are the electroweak precision tests telling us something relevant on
low-energy SUSY?
\begin{itemize}
  
\item[i)] SUSY must be a low-energy symmetry if it has to deal with
  the issue of the gauge hierarchy problem. This fact is usually
  translated into the statement that SUSY particle masses should not
  be significantly larger than $O(1~{\rm TeV})$ given that SUSY
  breaking should not exceed this energy scale to realize a suitable
  ``protection'' of the mass of the scalar Higgs responsible for the
  electroweak breaking. Actually one may try to be more quantitative
  \cite{barbgiu}. First one relates the $Z$ mass to the value of the 4
  parameters of the minimal SUGRA run from the large scale, at which
  the soft breaking terms originate, down to the electroweak scale.
  Then one establishes a degree of naturalness corresponding to the
  amount of fine tuning of the initial SUSY parameters which is needed
  to reproduce the correct $Z$ mass for increasing values of the
  low-energy SUSY masses. For instance it is clear that to have all
  SUSY particles with a mass of $O(1~{\rm TeV})$ would require a
  severe fine tuning of the boundary conditions.
  
  As for all naturalness criteria, also in this case there is a large
  amount of subjectivity, but one message emerges quite clearly:
  already now, in particular with the lower bound on chargino masses
  exceeding $90~{\rm GeV}$, we are entering an area of parameter space
  where a certain degree of fine tuning is needed. Hence we are
  already at the stage where we may expect ``naturally'' to find SUSY
  particles.  Moreover such naturalness analyses confirm that LHC
  represents kind of ``definitive'' machine for SUSY direct searches:
  if no SUSY particle is discovered at LHC, the degree of fine tuning
  becomes so severe that it is hard to still defend the idea of
  low-energy SUSY.  Finally, an important comment on the degree at
  which different SUSY masses are constrained by such naturalness
  criteria: due to the large difference in the Yukawa couplings of
  the third (heaviest) generation with respect to the first two
  generations, it turns out that only the sbottoms and stops are
  required not to be very heavy, whilst squarks of the first two
  generations can be quite heavy (say tens of TeV) without severely
  affecting the correct electroweak breaking. This observations may
  play a relevant role in tackling the FCNC problem in SUSY (see
  below).
 
\item[ii)] If one allows the SUSY parameters to take larger and larger
  values, all the SUSY particles become heavier and heavier with only
  one remarkable exception. In the Higgs mass spectrum of SUSY models
  the lightest scalar always remains light. The mass of the light
  CP-even neutral Higgs in the MSSM is calculable at tree level in
  terms of two SUSY parameters of the Higgs potential. At this level
  it is smaller than the mass of the $Z$. When radiative corrections
  are included, the mass of the light Higgs becomes a function also of
  the other SUSY parameters and its upper bound increases
  significantly \cite{zwirner}. However, even varying the MSSM
  parameters as much as one wishes, it is not possible to exceed
  $130-135~{\rm GeV}$ for its mass. Indeed, taking $m_t=175~{\rm GeV}$
  and for a stop lighter than $1~{\rm TeV}$ one obtains that the upper
  bound on the lightest Higgs is $125~{\rm GeV}$ allowing for
  ``maximal'' mixing in the top squark sector (the bound decreases for
  smaller stop mixing).
  
  It is not easy to significantly evade the above upper bound on the
  mass of the lightest Higgs even if one gives up the minimality of
  the SUSY model.  For instance, if one adds a singlet to the two
  Higgs doublets (i.e., one goes to the so called Next-to-Minimal SUSY
  Standard Model, NMSSM), then a new parameter shows up in the scalar
  potential: the coupling of the singlet with the two doublets. If one
  imposes that all couplings remain perturbative up to the Planck
  scale, then the consequent upper bound on this new coupling implies
  that the lightest Higgs should not be heavier than $150~{\rm GeV}$
  or so \cite{singlet}.
  
  Obviously having a possibly ``exotic'' Higgs below $150~{\rm GeV}$
  does not necessarily mean that it can be seen at LHC. While the
  lightest Higgs of the MSSM seems to be detectable at LHC, there may
  still be some significant loopholes for searches of the light Higgs
  in the NMSSM context.
  
\item[iii)] It is known that the MSSM is a decoupling theory. In the
  limit where we send the SUSY parameters to infinity all SUSY masses
  become infinite, with only the lightest Higgs remaining light and
  coinciding with the usual SM Higgs. In this limit we would recover
  the SM. It turns out that as far as electroweak precision tests are
  concerned, the decoupling of the MSSM is quite fast: already for
  SUSY masses above $200-300~{\rm GeV}$ the effects due to the
  exchange of SUSY particles in radiative corrections to the
  electroweak observables become negligible. Notice that this is not
  true if, instead of electroweak precision tests, we consider FCNC
  and CP tests. In this latter case, the decoupling may be much slower
  with squarks and gluinos of $1~{\rm TeV}$ still providing sizeable
  contributions in loop diagrams to some rare processes.
  
  Obviously the SM fit of electroweak precision data is now so good
  that there is no point in trying to improve it by the addition of
  the several degrees of freedom represented by the SUSY particles.
  The situation was different a couple of years ago when the
  discrepancy between the SM prediction and the data in the decay of
  the $Z$ into a $b$ quark pair resulted in a SM fit which could be
  significantly improved. Now the goal of the game has changed: one
  looks for regions of the SUSY parameter space where (some) SUSY
  masses are sufficiently small so that virtual SUSY contributions to
  electroweak observables are sizeable \cite{stefan}. Some of these
  regions may cause unbearably high departures from the SM predictions
  and hence they can be ruled out. In this way it is possible to
  exclude some (limited) portions of the MSSM parameter space which
  would be otherwise allowed by the limits on SUSY parameters coming
  from direct searches of SUSY particles.
   
  Finally, we make a comment related to the prediction of one
  low-energy parameter (the electroweak angle or, as it is the case
  nowadays, the value of the strong coupling at the $Z$ mass scale)
  when one asks for the unification of the gauge coupling constants in
  the MSSM. The value predicted for $\alpha_S(m_Z)$ in the MSSM is a
  couple of standard deviations higher than the experimental value. We
  do not consider this as a problem for the MSSM. Indeed, high energy
  thresholds generated from the masses of superheavy GUT particles may
  conceivably produce corrections able to account for such
  discrepancy. Taking into account the uncertainties in the dynamics
  at the GUT scale, we consider the argument of unification of
  couplings as a support to the existence of low-energy SUSY.
\end{itemize}

Before starting our discussion of indirect searches of SUSY, let us
emphasise that direct production and detection of SUSY particles
remain the only way to definitely prove the existence of low-energy
SUSY. However it is true that if LEP II is not going to find a SUSY
signal and unless some surprise possibly comes from Tevatron, we will
have to wait almost ten years to obtain an answer from such direct
searches. In view of this fact and of what we said in this section we
think that indirect searches of SUSY in the pre-LHC era deserve a very
special attention.

\section{FCNC and SUSY}
\label{sec:FCNC}

The generation of fermion masses and mixings (``flavour problem'')
gives rise to a first and important distinction among theories of new
physics beyond the electroweak standard model.

One may conceive a kind of new physics which is completely ``flavour
blind'', i.e. new interactions which have nothing to do with the
flavour structure. To provide an example of such a situation, consider
a scheme where flavour arises at a very large scale (for instance the
Planck mass) while new physics is represented by a supersymmetric
extension of the SM with supersymmetry broken at a much lower scale
and with the SUSY breaking transmitted to the observable sector by
flavour-blind gauge interactions \cite{GMSB1}-\cite{GMSB3}. In this
case one may think that new physics does not cause any major change to
the original flavour structure of the SM, namely that the pattern of
fermion masses and mixings is compatible with the numerous and
demanding tests of flavour changing neutral currents.

Alternatively, one can conceive a new physics which is entangled with
the flavour problem. As an example consider a technicolour scheme
where fermion masses and mixings arise through the exchange of new
gauge bosons which mix together ordinary and technifermions. Here we
expect (correctly enough) new physics to have potential problems in
accommodating the usual fermion spectrum with the adequate suppression
of FCNC. As another example of new physics which is not flavour blind,
take a more conventional SUSY model which is derived from a
spontaneously broken N=1 supergravity and where the SUSY breaking
information is conveyed to the ordinary sector of the theory through
gravitational interactions. In this case we may expect that the scale
at which flavour arises and the scale of SUSY breaking are not so
different and possibly the mechanism itself of SUSY breaking and
transmission is flavour-dependent. Under these circumstances we may
expect a potential flavour problem to arise, namely that SUSY
contributions to FCNC processes are too large.

The potentiality of probing SUSY in FCNC phenomena was readily
realized when the era of SUSY phenomenology started in the early 80's
\cite{SUSY2}.  In particular, the major implication that the scalar
partners of quarks of the same electric charge but belonging to
different generations had to share a remarkably high mass degeneracy
was emphasised.

Throughout the large amount of work in this last decade it became
clearer and clearer that generically talking of the implications of
low-energy SUSY on FCNC may be rather misleading. In minimal SUGRA
FCNC contributions can be computed in terms of a very limited set of
unknown new SUSY parameters. Remarkably enough, this minimal model
succeeds to pass all the set of FCNC tests unscathed. To be sure, it
is possible to severely constrain the SUSY parameter space, for
instance using $b \to s \gamma$, in a way which is complementary to
what is achieved by direct SUSY searches at colliders.

However, the MSSM is by no means equivalent to low-energy SUSY. A
first sharp distinction concerns the mechanism of SUSY breaking and
transmission to the observable sector which is chosen. As we mentioned
above, in models with gauge-mediated SUSY breaking (GMSB models
\cite{GMSB1}-\cite{GMSB3}) it may be possible to avoid the FCNC threat
``ab initio'' (notice that this is not an automatic feature of this
class of models, but it depends on the specific choice of the sector
which transmits the SUSY breaking information, the so-called messenger
sector). The other more ``canonical'' class of SUSY theories that was
mentioned above has gravitational messengers and a very large scale at
which SUSY breaking occurs. In this talk we will focus only on this
class of gravity-mediated SUSY breaking models. Even sticking to this
more limited choice we have a variety of options with very different
implications for the flavour problem.

First, there exists an interesting large class of SUSY realizations
where the customary R-parity (which is invoked to suppress proton
decay) is replaced by other discrete symmetries which allow either
baryon or lepton violating terms in the superpotential. But, even
sticking to the more orthodox view of imposing R-parity, we are still
left with a large variety of extensions of the MSSM at low energy. The
point is that low-energy SUSY ``feels'' the new physics at the
superlarge scale at which supergravity (i.e., local supersymmetry)
broke down. In this last couple of years we have witnessed an
increasing interest in supergravity realizations without the so-called
flavour universality of the terms which break SUSY explicitly. Another
class of low-energy SUSY realizations which differ from the MSSM in
the FCNC sector is obtained from SUSY-GUT's. The interactions
involving superheavy particles in the energy range between the GUT and
the Planck scale bear important implications for the amount and kind
of FCNC that we expect at low energy.

Given a specific SUSY model it is in principle possible to make a full
computation of all the FCNC phenomena in that context. However, given
the variety of options for low-energy SUSY (even confining ourselves
here to models with R matter parity), it is important to have a way to
extract from the whole host of FCNC processes a set of upper limits on
quantities which can be readily computed in any chosen SUSY frame.

The best model-independent parameterisation of FCNC effects is the
so-called mass insertion approximation \cite{mins}.  It concerns the
most peculiar source of FCNC SUSY contributions that do not arise from
the mere supersymmetrization of the FCNC in the SM. They originate
from the FC couplings of gluinos and neutralinos to fermions and
sfermions~\cite{FCNC}. One chooses a basis for the fermion and
sfermion states where all the couplings of these particles to neutral
gauginos are flavour diagonal, while the FC is exhibited by the
non-diagonality of the sfermion propagators. Denoting by $\Delta$ the
off-diagonal terms in the sfermion mass matrices (i.e. the mass terms
relating sfermion of the same electric charge, but different flavour),
the sfermion propagators can be expanded as a series in terms of
$\delta = \Delta/ \tilde{m}^2$, where $\tilde{m}$ is the average
sfermion mass.  As long as $\Delta$ is significantly smaller than
$\tilde{m}^2$, we can just take the first term of this expansion and,
then, the experimental information concerning FCNC and CP violating
phenomena translates into upper bounds on these $\delta$'s
\cite{deltas}-\cite{GGMS}.

Obviously the above mass insertion method presents the major advantage
that one does not need the full diagonalisation of the sfermion mass
matrices to perform a test of the SUSY model under consideration in
the FCNC sector. It is enough to compute ratios of the off-diagonal
over the diagonal entries of the sfermion mass matrices and compare
the results with the general bounds on the $\delta$'s that we provide
here from all available experimental information.

There exist four different $\Delta$ mass insertions connecting
flavours $i$ and $j$ along a sfermion propagator:
$\left(\Delta_{ij}\right)_{LL}$, $\left(\Delta_{ij}\right)_{RR}$,
$\left(\Delta_{ij}\right)_{LR}$ and $\left(\Delta_{ij}\right)_{RL}$.
The indices $L$ and $R$ refer to the helicity of the fermion partners.
Instead of the dimensional quantities $\Delta$ it is more useful to
provide bounds making use of dimensionless quantities, $\delta$, that
are obtained dividing the mass insertions by an average sfermion mass.

Let us first consider CP-conserving $\Delta F=2$ processes.  The
amplitudes for gluino-mediated contributions to $\Delta F=2$
transitions in the mass-insertion approximation have been computed in
refs.~\cite{GMS,GGMS}. Imposing that the contribution to $K-\bar K$,
$D-\bar D$ and $B_d - \bar{B}_d$ mixing proportional to each single
$\delta$ parameter does not exceed the experimental value, we obtain
the constraints on the $\delta$'s reported in table~\ref{reds2},
barring accidental cancellations \cite{GGMS} (for a QCD-improved
computation of the constraints coming from $K-\bar K$ mixing, see
ref.~\cite{bagger}).

\begin{table}
 \begin{center}
 \begin{tabular}{||c|c|c|c||}  \hline \hline
 $x$ & $\sqrt{\left|\Re  \left(\delta^{d}_{12} \right)_{LL}^{2}\right|} $ 
 &
 $\sqrt{\left|\Re  \left(\delta^{d}_{12} \right)_{LR}^{2}\right|} $ &
 $\sqrt{\left|\Re  \left(\delta^{d}_{12} \right)_{LL}\left(\delta^{d}_{12}
 \right)_{RR}\right|} $ \\
 \hline
 $
   0.3
 $ &
 $
1.9\times 10^{-2}
 $ & $
7.9\times 10^{-3}
 $ & $
2.5\times 10^{-3}
 $ \\
 $
   1.0
 $ &
 $
4.0\times 10^{-2}
 $ & $
4.4\times 10^{-3}
 $ & $
2.8\times 10^{-3}
 $ \\
 $
   4.0
 $ &
 $
9.3\times 10^{-2}
 $ & $
5.3\times 10^{-3}
 $ & $
4.0\times 10^{-3}
 $ \\ \hline \hline
 $x$ & $\sqrt{\left|\Re  \left(\delta^{d}_{13} \right)_{LL}^{2}\right|} $ 
 &
 $\sqrt{\left|\Re  \left(\delta^{d}_{13} \right)_{LR}^{2}\right|} $ &
 $\sqrt{\left|\Re  \left(\delta^{d}_{13} \right)_{LL}\left(\delta^{d}_{13}
 \right)_{RR}\right|} $ \\
 \hline
 $
   0.3
 $ &
 $
4.6\times 10^{-2}
 $ & $
5.6\times 10^{-2}
 $ & $
1.6\times 10^{-2}
 $ \\
 $
   1.0
 $ &
 $ 
9.8\times 10^{-2}
 $ & $
3.3\times 10^{-2}
 $ & $
1.8\times 10^{-2}
 $ \\
 $
   4.0
 $ &
 $
2.3\times 10^{-1}
 $ & $
3.6\times 10^{-2}
 $ & $
2.5\times 10^{-2}
 $ \\ \hline \hline
 $x$ & $\sqrt{\left|\Re  \left(\delta^{u}_{12} \right)_{LL}^{2}\right|} $ 
 &
 $\sqrt{\left|\Re  \left(\delta^{u}_{12} \right)_{LR}^{2}\right|} $ &
 $\sqrt{\left|\Re  \left(\delta^{u}_{12} \right)_{LL}\left(\delta^{u}_{12}
 \right)_{RR}\right|} $ \\
 \hline
 $
   0.3
 $ &
 $
4.7\times 10^{-2}
 $ & $
6.3\times 10^{-2}
 $ & $
1.6\times 10^{-2}
 $ \\
 $
   1.0
 $ &
 $
1.0\times 10^{-1}
 $ & $
3.1\times 10^{-2}
 $ & $
1.7\times 10^{-2}
 $ \\
 $
   4.0
 $ &
 $
2.4\times 10^{-1}
 $ & $
3.5\times 10^{-2}
 $ & $
2.5\times 10^{-2}
 $ \\ \hline \hline
\end{tabular}
\caption[]{Limits on $\mbox{Re}\left(\delta_{ij}\right)_{AB}\left(
     \delta_{ij}\right)_{CD}$, with $A,B,C,D=(L,R)$, for an average
   squark mass $m_{\tilde{q}}=500\mbox{GeV}$ and for different values
   of $x=m_{\tilde{g}}^2/m_{\tilde{q}}^2$. For different values of
   $m_{\tilde{q}}$, the limits can be obtained multiplying the ones in
   the table by $m_{\tilde{q}}(\mbox{GeV})/500$.}
 \label{reds2}
\end{center}
\end{table} 
 
We then consider the process $b \to s \gamma$. This decay requires a
helicity flip. In the presence of a $\left(\delta^d_{23}\right)_{LR}$
mass insertion we can realize this flip in the gluino running in the
loop. On the contrary, the $\left( \delta^d_{23}\right)_{LL}$
insertion requires the helicity flip to occur in the external
$b$-quark line. Hence we expect a stronger bound on the
$\left(\delta^d_{23}\right)_{LR}$ quantity. Indeed, this is what
happens: $\left(\delta^d_{23}\right)_{LL}$ is essentially not bounded,
while $\left(\delta^d_{23}\right)_{LR}$ is limited to be
$<10^{-3}-10^{-2}$ according to the average squark and gluino masses
\cite{GGMS}.

Given the upper bound on $\left(\delta^d_{23}\right)_{LR}$ from $b \to
s \gamma$, it turns out that the quantity $x_{s}$ of the
$B_{s}-\bar{B}_{s}$ mixing receives contributions from this kind of
mass insertions which are very tiny. The only chance to obtain large
values of $x_s$ is if $\left(\delta^d_{23}\right)_{LL}$ is large, say
of $O(1)$. In that case $x_s$ can easily jump up to values of $O
(10^{2})$ or even larger.

Then, imposing the bounds in table~\ref{reds2}, we can obtain the
largest possible value for BR($b \to d \gamma$) through gluino
exchange. As expected, the $\left( \delta^{d}_{13}\right)_{LL}$
insertion leads to very small values of this BR of $O(10^{-7})$ or so,
whilst the $\left( \delta^{d}_{13}\right)_{LR}$ insertion allows for
BR($b \to d \gamma$) ranging from few times $10^{-4}$ up to few times
$10^{-3}$ for decreasing values of $x=m^{2}_{\tilde{g}}/
m^{2}_{\tilde{q}}$.  In the SM we expect BR($b \to d \gamma$) to be
typically $10-20$ times smaller than BR($b \to s \gamma$), i.e. BR($b
\to d \gamma)=(1.7\pm 0.85 )\times 10^{-5}$. Hence a large enhancement
in the SUSY case is conceivable if $\left(
  \delta^{d}_{13}\right)_{LR}$ is in the $10^{-2}$ range. Notice that
in the MSSM we expect $\left( \delta^{d}_{13}\right)_{LR}<m^{2}_{b}/
m^{2}_{\tilde{q}}\times V_{td}<10^{-6}$, hence with no hope at all of
a sizeable contribution to $b \to d \gamma$.

An analysis similar to the one of $b \to s \gamma$ decays can be
performed in the leptonic sector where the masses $m_{\tilde{q}}$ and
$m_{\tilde{g}}$ are replaced by the average slepton mass
$m_{\tilde{l}}$ and the photino mass $m_{\tilde{\gamma}}$
respectively.  The most stringent bound concerns the transition $\mu
\to e \gamma$ with $\left(\delta^l_{12}\right)_{LR}<10^{-6}$ for
slepton and photino masses of O(100 GeV) \cite{GGMS}.

\section{CP and SUSY}
\label{sec:CP}

The situation concerning CP violation in the MSSM case with
$\Phi_A=\Phi_B=0$ and exact universality in the soft-breaking sector
can be summarised in the following way: the MSSM does not lead to any
significant deviation from the SM expectation for CP-violating
phenomena as $d_N^e$, $\varepsilon$, $\varepsilon^\prime$ and CP
violation in $B$ physics; the only exception to this statement
concerns a small portion of the MSSM parameter space where a very
light $\tilde t$ ($m_{\tilde t} < 100$ GeV) and $\chi^+$ ($m_\chi \sim
90$ GeV) are present. In this latter particular situation sizeable
SUSY contributions to $\varepsilon_K$ are possible and, consequently,
major restrictions in the $\rho-\eta$ plane can be inferred (see, for
instance, ref. \cite{misiakpok}). Obviously, CP violation in $B$
physics becomes a crucial test for this MSSM case with very light
$\tilde t$ and $\chi^+$. Interestingly enough, such low values of SUSY
masses are at the border of the detectability region at LEP II.

We now turn to CP violation in the model-independent approach that we
are proposing here. For a detailed discussion we refer the reader to
our general study \cite{GGMS}. Here we just summarise the situation in
the following three points:
\begin{itemize}
\item[i)] $\epsilon$ provides bounds on the imaginary parts of the
  quantities whose real part was limited by the $K$ mass difference
  which are roughly one order of magnitude more severe than the
  corresponding ones derived from $\Delta m_K$.
  
\item[ii)] The nature of the SUSY contribution to CP violation is
  generally superweak, since the constraints from $\varepsilon$ are
  always stronger (in the left-left sector) or at least equal (in the
  left-right sector) to the ones coming from
  $\varepsilon^\prime/\varepsilon$.
  
\item[iii)] the experimental bound on the electric dipole moment of
  the neutron imposes very stringent limits on ${\rm
    Im}\left(\delta_{11}^{d}\right)_{LR}$ (of $O(10^{-6})$ for an
  average squark and gluino mass of $500~{\rm GeV}$.)  In conclusion,
  although technically it is conceivable that some SUSY extension may
  provide a sizable contribution to
  $\varepsilon^{\prime}/\varepsilon$, it is rather difficult to
  imagine how to reconcile a relatively large value of ${\rm
    Im}\left(\delta_{12}^{d}\right)_{LR}$ with the very strong
  constraint on the flavour-conserving ${\rm
    Im}\left(\delta_{11}^{d}\right)_{LR}$ from $d^e_N$.
\end{itemize}

We now move to the next frontier for testing the unitarity triangle in
general and in particular CP violation in the SM and its SUSY
extensions: $B$ physics. We have seen above that the transitions
between 1st and 2nd generation in the down sector put severe
constraints on $\Re \delta^d_{12}$ and $\Im \delta^d_{12}$ quantities.
To be sure, the bounds derived from $\varepsilon$ and
$\varepsilon^\prime$ are stronger than the corresponding bounds from
$\Delta M_K$. If the same pattern repeats itself in the transition
between 3rd and 1st or 3rd and 2nd generation in the down sector we
may expect that the constraints inferred from $B_d - \bar{B}_d$
oscillations or $b \to s \gamma$ do not prevent conspicuous new
contributions also in CP violating processes in $B$ physics. We are
going to see below that this is indeed the case ad we will argue that
measurements of CP asymmetries in several $B$-decay channels may allow
to disentangle SM and SUSY contributions to the CP decay phase.

New physics can modify the SM predictions on CP asymmetries in $B$
decays by changing the phase of the $B_{d}$--$\bar{B}_{d}$ mixing and
the phase and absolute value of the decay amplitude.  The general SUSY
extension of the SM that we discuss here affects both these
quantities.

The crucial question is then: where and how can one possibly
distinguish SUSY contributions to CP violation in $B$ decays
\cite{cptutti}?

In terms of the decay amplitude $A$, the CP asymmetry reads 
\begin{equation}
{\cal A}(t) = \frac{(1-\vert \lambda\vert^2) \cos (\Delta M_d t )
-2 {\rm Im} \lambda \sin (\Delta M_d t )}{1+\vert \lambda\vert^2} 
\label{eq:asy}
\end{equation}
with $\lambda=e^{-2i\phi^M}\bar{A}/A$. 
In order to be able to discuss the results model-independently, we
have labeled as $\phi^M$ the generic mixing phase.  The ideal case
occurs when one decay amplitude only appears in (or dominates) a decay
process: the CP violating asymmetry is then determined by the total
phase $\phi^T=\phi^M+\phi^D$, where $\phi^D$ is the weak phase of the
decay.  This ideal situation is spoiled by the presence of several
interfering amplitudes.

We summarise the results in table~\ref{tab:results} which is taken
from the recent analysis of ref.~\cite{cpnoi}. We refer the interested
reader to our work \cite{cpnoi} for all the details of how our
computation in the SM and in SUSY is carried out.  $\Phi^D_{SM}$
denotes the decay phase in the SM; for each channel, when two
amplitudes with different weak phases are present, we indicate the SM
phase of the Penguin (P) and Tree-level (T) decay amplitudes.  For $B
\to K_S \pi^{0}$ the penguin contributions (with a vanishing phase)
dominate over the tree-level amplitude because the latter is Cabibbo
suppressed.  For the channel $b \to s \bar s d$ only penguin operators
or penguin contractions of current-current operators contribute. The
phase $\gamma$ is present in the penguin contractions of the $(\bar b
u)(\bar u d)$ operator, denoted as $u$-P $\gamma$ in
table~\ref{tab:results}.  $\bar b d \to \bar q q $ indicates processes
occurring via annihilation diagrams which can be measured from the
last two channels of table~\ref{tab:results}.  In the case $B \to
K^{+} K^{-}$ both current-current and penguin operators contribute. In
$B \to D^{0} \bar D^{0}$ the contributions from the $(\bar b u) (\bar
u d)$ and the $(\bar b c) (\bar c d)$ current-current operators
(proportional to the phase $\gamma$) tend to cancel out.

\begin{table}
 \begin{center}
 \begin{tabular}{|ccccccc|}
 \hline 
 \ss{Incl. }&\ss{ Excl. }&\ss{ $\phi^{D}_{\rm SM}$ }&\ss{ $r_{\rm SM}$ }&\ss{ 
 $\phi^{D}_{\rm SUSY}$ }&\ss{ $r_{250}$ }&\ss{ $r_{500}$ }\\ 
 \hline
\ss{ $b \to c \bar c s$ }&\ss{ $B \to J/\psi K_{S}$ }&\ss{ 0 }&\ss{ -- }&\ss{
 $\phi_{23}$ }&\ss{ $0.03-0.1$ 
 }&\ss{$0.008-0.04$ }\\ \hline
 \ss{ $b \to s \bar s s$ }&\ss{ $B \to \phi K_{S}$ }&\ss{ 0 }&\ss{ -- }&\ss{
 $\phi_{23}$ }&\ss{ $0.4-0.7$ }&\ss{ 
 $0.09-0.2$ }\\ \hline \ss{
 $b \to u \bar u s$ }&\ss{ }&\ss{ P $0$ }&\ss{  }&\ss{  }&\ss{  }&\ss{
  }\\ 
\ss{}&\ss{$ B \to \pi^{0} K_{S}$} &\ss{  }&\ss{ $0.01-0.08$ }&\ss{
  $\phi_{23}$ }&\ss{ $0.4-0.7$ }&\ss{ 
 $0.09-0.2$ }\\ \ss{ 
 $b \to d \bar d s$ }&\ss{ }&\ss{ T $\gamma$ }&\ss{  }&\ss{  }&\ss{  }&\ss{
  }\\ \hline \ss{
 $b \to c \bar u d$ }&\ss{ }&\ss{ 0 }&\ss{  }&\ss{  }&\ss{  }&\ss{
  }\\ \ss{ 
  }&\ss{$ B \to D^{0}_{CP} \pi^{0}$ }&\ss{  }&\ss{ 0.02 }&\ss{ -- }&
  \ss{ -- }&\ss{
 -- }\\ \ss{  
 $b \to u \bar c d$ }&\ss{ }&\ss{ $\gamma$ }&\ss{  }&\ss{  }&\ss{  }&\ss{
  }\\ \hline \ss{
  }&\ss{ $B \to D^{+} D^{-}$ }&\ss{ T $0$ }&\ss{ $0.03-0.3$ }&\ss{  }&\ss{
  $0.007-0.02$ }&\ss{ 
  $0.002-0.006$ }\\ \ss{ 
  $b \to c \bar c d$}&\ss{ }&\ss{ }&\ss{  }&\ss{ $\phi_{13}$ }&\ss{ }&\ss{
  }\\ \ss{ 
  }&\ss{ $B \to J/\psi \pi^{0}$ }&\ss{ P $\beta$ }&\ss{ $0.04-0.3$ }&\ss{
  }&\ss{ $0.007-0.03$ }&\ss{ 
  $0.002-0.008$ 
  }\\ \hline \ss{
  }&\ss{ $B \to \phi \pi^{0}$ }&\ss{ P $\beta$}&\ss{ -- }&\ss{ }&\ss{
  $0.06-0.1$ }&\ss{ 
  $0.01-0.03$ }\\ \ss{ 
  $b \to s \bar s d$}&\ss{ }&\ss{  } &&\ss{ $\phi_{13}$ }&\ss{ }&\ss{
  }\\ \ss{ 
  }&\ss{ $B \to K^{0} \bar{K}^{0}$ }&\ss{ {\it u}-P
  $\gamma$  
  }&\ss{ $0-0.07$ }&\ss{ }&\ss{ $0.08-0.2$ }&\ss{
  $0.02-0.06$ 
  }\\ \hline \ss{
 $b \to u \bar u d$ }&\ss{ $B \to \pi^{+} \pi^{-}$ }&\ss{ T
 $\gamma$  
 }&\ss{ $0.09-0.9$ }&\ss{ $\phi_{13}$ }&\ss{ $0.02-0.8$ }&\ss{
 $0.005-0.2$ }\\ \ss{ 
 $b \to d \bar d d$ }&\ss{ $B \to \pi^{0} \pi^{0}$ }&\ss{
 P $\beta$ }&\ss{ $0.6-6$  
 }&\ss{ $\phi_{13}$ }&\ss{ $0.06-0.4$ }&\ss{
 $0.02-0.1$ }\\ \hline \ss{
 }&\ss{ $B \to K^{+} K^{-}$ }&\ss{ T $\gamma$ }&\ss{ $0.2-0.4$ }&\ss{ }&\ss{
 $0.04-0.1$ }&\ss{$0.01-0.03$ 
  }\\ \ss{
 $b \bar d \to q \bar q$ }&\ss{ }&\ss{ }&\ss{ }&\ss{ $\phi_{13}$}&
 \ss{ }&\ss{ }\\ \ss{
 }&\ss{ $B \to D^{0} \bar D^{0}$ }&\ss{ P $\beta$ 
 }&\ss{ only $\beta$ }&\ss{  }&\ss{ $0.01-0.03$ }&\ss{$0.003-0.006$ }\\  \hline
\end{tabular}
\caption[]{CP phases for $B$ decays. $\phi^{D}_{SM}$
  denotes the decay phase in the SM; T and P denote Tree and Penguin,
  respectively; for each channel, when two amplitudes with different
  weak phases are present, one is given in the first row, the other in
  the last one and the ratio of the two in the $r_{SM}$ column.
  $\phi^{D}_{SUSY}$ denotes the phase of the SUSY amplitude, and the
  ratio of the SUSY to SM contributions is given in the $r_{250}$ and
  $r_{500}$ columns for the corresponding SUSY masses.}
 \label{tab:results}
\end{center}
\end{table}

SUSY contributes to the decay amplitudes with phases induced by
$\delta_{13}$ and $\delta_{23}$ which we denote as $\phi_{13}$ and
$\phi_{23}$. The ratios of $A_{SUSY}/A_{SM}$ for SUSY masses of 250
and 500 GeV are reported in the $r_{250}$ and $r_{500}$ columns of
table~\ref{tab:results}.

We now draw some conclusions from the results of
table~\ref{tab:results}.  In the SM, the first six decays measure
directly the mixing phase $\beta$, up to corrections which, in most of
the cases, are expected to be small.  These corrections, due to the
presence of two amplitudes contributing with different phases, produce
uncertainties of $\sim 10$\% in $B \to K_S \pi^{0}$, and of $\sim
30$\% in $B \to D^{+} D^{-}$ and $B \to J/\psi \pi^{0}$.  In spite of
the uncertainties, however, there are cases where the SUSY
contribution gives rise to significant changes.  For example, for SUSY
masses of O(250) GeV, SUSY corrections can shift the measured value of
the sine of the phase in $B \to \phi K_S$ and in $B \to K_S \pi^{0}$
decays by an amount of about 70\%.  For these decays SUSY effects are
sizeable even for masses of 500 GeV.  In $B \to J/\psi K_S$ and $B \to
\phi \pi^0$ decays, SUSY effects are only about $10$\% but SM
uncertainties are negligible.  In $B \to K^0 \bar{K}^0$ the larger
effect, $\sim 20$\%, is partially covered by the indetermination of
about $10$\% already existing in the SM.  Moreover the rate for this
channel is expected to be rather small.  In $B \to D^{+} D^{-}$ and $B
\to K^{+} K^{-}$, SUSY effects are completely obscured by the errors
in the estimates of the SM amplitudes.  In $B^0\to D^0_{CP}\pi^0$ the
asymmetry is sensitive to the mixing angle $\phi_M$ only because the
decay amplitude is unaffected by SUSY.  This result can be used in
connection with $B^0 \to K_s \pi^0$, since a difference in the measure
of the phase is a manifestation of SUSY effects.

Turning to $B \to \pi \pi$ decays, both the uncertainties in the SM
and the SUSY contributions are very large. Here we witness the
presence of three independent amplitudes with different phases and of
comparable size.  The observation of SUSY effects in the $\pi^{0}
\pi^{0}$ case is hopeless. The possibility of separating SM and SUSY
contributions by using the isospin analysis remains an open
possibility which deserves further investigation.  For a thorough
discussion of the SM uncertainties in $B \to \pi \pi $ see
ref.~\cite{cfms}.

In conclusion, our analysis shows that measurements of CP asymmetries
in several channels may allow the extraction of the CP mixing phase
and to disentangle SM and SUSY contributions to the CP decay phase.
The golden-plated decays in this respect are $B \to \phi K_S$ and $B
\to K_S \pi^0$ channels. The size of the SUSY effects is clearly
controlled by the the non-diagonal SUSY mass insertions $\delta_{ij}$,
which for illustration we have assumed to have the maximal value
compatible with the present experimental limits on $B^0_d$--$\bar
B^0_d$ mixing.

\section{DM and SUSY: a brief comment}

We have strong indications that ordinary matter (baryons) is
insufficient to provide the large amount of non-shining matter which
has been experimentally proven to exist in galactic halos and at the
level of clusters of galaxies \cite{kolb}. In a sense, this might
constitute the ``largest'' indication of new physics beyond the SM.
This statement holds true even after the recent stunning developments
in the search for non-shining baryonic objects.  In September 1993 the
discovery of massive dark objects (``machos'') was announced. After
five years of intensive analysis it is now clear that in any case
machos cannot account for the whole dark matter of the galactic halos.

It was widely expected that some amount of non-shining baryonic matter
could exist given that the contribution of luminous baryons to the
energy density of the Universe $\Omega$ = $\rho/\rho_{cr}$
($\rho_{cr}= 3H^2_0 /8 \pi G$ where G is the gravitational constant
and $H_0$ the Hubble constant) is less than $1\%$, while from
nucleosynthesis we infer $\Omega_{baryon} = \rho_{baryon} / \rho_{cr}
= (0.06\- \pm 0.02) h_{50}^{-2}$, where $h_{50} = H_0 /50$ Km/s Mpc.
On the other hand, we have direct indications that $\Omega$ should be
at least $20\%$ which means that baryons can represent not more than
half of the entire energy density of the Universe \cite{kolb}.

We could make these considerations on the insufficiency of the SM to
obtain a large enough $\Omega$ more dramatic if we accept the
theoretical input that the Universe underwent some inflationary era
which produced $\Omega$ =$1$. In that case, at least $90\%$ of the
whole energy density of the Universe should be provided by some new
physics beyond the SM.

Before discussing possible particle physics candidates, it should be
kept in mind that DM is not only called for to provide a major
contribution to $\Omega$, but also it has to provide a suitable
gravitational driving force for the primordial density fluctuations to
evolve into the large-scale structures (galaxies, clusters and
superclusters of galaxies) that we observe today \cite{kolb}. Here we
encounter the major difficulties when dealing with the two
``traditional'' sources of DM: Cold (CDM) and Hot (HDM) DM.

Light neutrinos in the eV range are the most typical example of HDM,
being their decoupling temperature of O(1 MeV). On the other hand, the
Lightest Supersymmetric Particle (LSP) in the tens of GeV range is a
typical CDM candidate. Taking the LSP to be the lightest neutralino,
one obtains that when it decouples it is already non-relativistic,
being its decoupling temperature typically one order of magnitude
below its mass.

Both HDM and CDM have some difficulty to correctly reproduce the
experimental spectrum related to the distribution of structures at
different scales. The conflict is more violent in the case of pure
HDM.  Neutrinos of few eV's tend to produce too many superlarge
structures. The opposite problem arises with pure CDM: we obtain too
much power in the spectrum at low mass scales (galactic scales).

A general feature is that some amount of CDM should be present in any
case.  A possibility which has been envisaged is that after all the
whole $\Omega$ could be much smaller than one, say $20\%$ or so and
then entirely due to CDM. However, if one keeps on demanding the
presence of an inflationary epoch, then it seems unnatural to have
$\Omega$ so different from unity (although lately some variants of
inflationary schemes leading to $\Omega$ smaller than one have been
proposed).  Another possibility is that CDM provides its $20\%$ to
$\Omega$, while all the rest to reach the unity value is given by a
nonvanishing cosmological constant.

Finally, the possibility which encounters quite some interest is the
so-called Mixed Dark Matter (MDM) \cite{shafi}, where a wise cocktail
of HDM and CDM is present. An obvious realization of a MDM scheme is a
variant of the MSSM where neutrinos get a mass of few eV's. In that
case the lightest neutralino (which is taken to be the LSP) plays the
role of CDM and the light neutrino(s) that of HDM. With an appropriate
choice of the parameters it is possible to obtain contributions to
$\Omega$ from the CDM and HDM in the desired range.

In the MSSM with R parity the lightest SUSY particle (LSP) is
absolutely stable. For several reasons the lightest neutralino is the
fa\-vourite candidate to be the LSP fulfilling the role of CDM
\cite{jungman}.

The neutralinos are the eigenvectors of the mass matrix of the four
neutral fermions partners of the $W_3, B, H^0_1$ and $H^0_2$. There
are four parameters entering this matrix: $M_1, M_2, \mu$ and $\tan
\beta$.  The first two parameters denote the coefficients of the SUSY
breaking mass terms $\tilde B \tilde B$ and $\tilde W_3 \tilde W_3$
respectively.  $\mu$ is the coupling of the $H_1H_2$ term in the
superpotential.  Finally $\tan \beta$ denotes the ratio of the VEV's
of the $H_2$ and $H_1$ scalar fields.

In general $M_1$ and $M_2$ are two independent parameters, but if one
assumes that grand unification takes place, then at the grand
unification scale $M_1 = M_2 = M_3$, where $M_3$ is the gluino mass at
that scale. Then at $M_W$ one obtains:
\begin{equation}
M_1 = {5 \over 3} \tan^2 \theta_w M_2 \simeq {M_2 \over 2}, \qquad
M_2 = {g^2_2 \over g^2_3} m_{\tilde g} \simeq {m_{\tilde g} \over 3},
\end{equation}
where $g_2$ and $g_3$ are the SU(2) and SU(3) gauge coupling constants, 
respectively.

The above relation between $M_1$ and $M_2$ reduces to three the number
of independent parameters which determine the lightest neutralino
composition and mass: $\tan \beta, \mu$ and $M_2$.  Hence, for fixed
values of $\tan \beta$ one can study the neutralino spectrum in the
($\mu, M_2$) plane. The major experimental inputs to exclude regions
in this plane are the request that the lightest chargino be heavier
than $M_Z$ and the limits on the invisible width of the Z hence
limiting the possible decays $Z \rightarrow \chi \chi,\-\chi \chi'$.

Let us focus now on the role played by $\chi$ as a source of CDM.
$\chi$ is kept in thermal equilibrium through its electroweak
interactions not only for $T > m_\chi$, but even when T is below
$m_\chi$. However for $T < m_\chi$ the number of $\chi's$ rapidly
decrease because of the appearance of the typical Boltzmann
suppression factor exp ($- m_\chi /T$). When T is roughly $m_\chi /20$
the number of $\chi$ diminished so much that they do not interact any
longer, i.e. they decouple.  Hence the contribution to $\Omega_{CDM}$
of $\chi$ is determined by two parameters: $m_\chi$ and the
temperature at which $\chi$ decouples $(T_D)$. $T_D$ fixes the number
of $\chi's$ which survive. As for the determination of $T_D$ itself,
one has to compute the $\chi$ annihilation rate and compare it with
the cosmic expansion rate \cite{ellis}.

Several annihilation channels are possible with the exchange of
different SUSY or ordinary particles, $\tilde f$, $H$, $Z$, etc.
Obviously the relative importance of the channels depends on the
composition of $\chi$. For instance, if $\chi$ is a pure gaugino, then
the $\tilde f$ exchange represents the dominant annihilation mode.

Quantitatively \cite{bottino}, it turns out that if $\chi$ results
from a large mixing of the gaugino ($\tilde W_3$ and $\tilde B$) and
Higgsino ($\tilde H^0_1$ and $\tilde H^0_2$) components, then the
annihilation is too efficient to allow the surviving $\chi$ to provide
$\Omega$ large enough. Typically in this case $\Omega < 10^{-2}$ and
hence $\chi$ is not a good CDM candidate. On the contrary, if $\chi$
is either almost a pure Higgsino or a pure gaugino then it can give a
conspicuous contribution to $\Omega$

In the case $\chi$ is mainly a gaugino (say at least at the $90 \%$
level) what is decisive to establish the annihilation rate is the mass
of $\tilde f$.  If sfermions are light the $\chi$ annihilation rate is
fast and the $\Omega_\chi$ is negligible.  On the other hand, if
$\tilde f$ (and hence $\tilde l$, in particular) is heavier than 150
GeV, the annihilation rate of $\chi$ is sufficiently suppressed so
that $\Omega_\chi$ can be in the right ballpark for $\Omega_{CDM}$. In
fact if all the $\tilde f's$ are heavy, say above 500 GeV and for
$m_\chi \ll m_{\tilde f}$, then the suppression of the annihilation
rate can become even too efficient yielding $\Omega_\chi$ unacceptably
large.

In the minimal SUSY standard model there are five new parameters in
addition to those already present in the non--SUSY case.  Imposing the
electroweak radiative breaking further reduces this number to four.
Finally, in simple supergravity realizations the soft parameters A and
B are related. Hence we end up with only three new, independent
parameters. One can use the constraint that the relic $\chi$ abundance
provides a correct $\Omega_{CDM}$ to restrict the allowed area in this
3--dimensional space.  Or, at least, one can eliminate points of this
space which would lead to $\Omega_\chi >1$, hence overclosing the
Universe. For $\chi$ masses up to 150 GeV it is possible to find
sizable regions in the SUSY parameter space where $\Omega_\chi$
acquires interesting values for the DM problem. A detailed discussion
on this point is beyond the scope of this talk. The interested reader
can find a thorough analysis in the review of Ref.~\cite{jungman} and
the original papers therein quoted.

There exist two ways to search for the existence of relic neutralinos.
First we have direct detection: neutralinos interact with matter both
through coherent and spin dependent effects. Only coherent effects are
currently accessible to direct detection. The sensitivity of the
direct detection experiment has reached now an area of the SUSY
parameter space of the MSSM which is of great interest for neutralinos
in the 50 GeV - 200 GeV range.

The indirect detection is based on the search for signals coming from
pair annihilation of neutralinos. Such annihilation may occur inside
celestial bodies (Earth, Sun, etc.) where neutralinos may be
gravitationally captured. The signal is then a flux of muon neutrinos
which can be detected as up-going muons in a neutrino telescope.
Another possibility is that the neutralino annihilation occurs in the
galactic halo. In this case the signal consists of photon, positron
and antiproton fluxes. They can be observed by detectors placed on
balloons or satellites. The computation of these fluxes is strongly
affected by the composition of the lightest neutralino. In any case
also these indirect searches for relic neutralinos are now probing
interesting areas of the MSSM parameter space.

A very different prospect for DM occurs in the GMSB schemes.  In this
case the gravitino mass ($m_{3/2}$) loses its role of fixing the
typical size of soft breaking terms and we expect it to be much
smaller than what we have in models with a hidden sector.  Indeed,
given the well-known relation \cite{SUSY1} between $m_{3/2}$ and the
scale of SUSY breaking $\sqrt{F}$, i.e.\ $m_{3/2}=O(F/M)$, where $M$
is the reduced Planck scale, we expect $m_{3/2}$ in the KeV range for
a scale $\sqrt{F}$ of $O(10^6$ GeV) that has been proposed in models
with low-energy SUSY breaking in a visible sector.

A gravitino of that mass behaves as a Warm Dark Matter (WDM) particle,
that is, a particle whose free streaming scale involves a mass
comparable to that of a galaxy, $\sim 10^{11-12}M_\odot$.

However, critical density models with pure WDM are known to suffer for
serious troubles \cite{colo}. Indeed, a WDM scenario behaves much like
CDM on scales above $\lambda_{FS}$. Therefore, we expect in the light
gravitino scenario that the level of cosmological density fluctuations
on the scale of galaxy clusters ($\sim 10\,h^{-1}$ Mpc) to be almost
the same as in CDM. As a consequence, the resulting number density of
galaxy clusters is predicted to be much larger than what observed.

We have recently considered different variants of a light gravitino DM
dominated model. It seems that in all cases there exist difficulties to
account correctly for cosmic straucture formation. This provides severe
cosmological constraints on the GMSB models \cite{pierpa}. 

In conclusion SUGRA models with R parity offer the best candidate for CDM.
It is remarkable that as a by-product of the MSSM we obtain a lightest
neutralino which can provide the correct amount of DM in a wide area of
the SUSY parametr space. Even more interesting, we are now experimentally
approaching the level of sensitivity which is needed to explore
(directly or indirectly) large portions of this area of parameter space.
The complementarity of this exploration to that performed by using FCNC and
CP tests and direct collider SUSY searches  looks promising.

\section*{Acknowledgements}

We are grateful to our ``FCNC collaborators'' M. Ciuchini, E. Franco,
F.  Gabbiani, E. Gabrielli and G. Martinelli and our ``DM
collaborators'' S.  Borgani, E. Pierpaoli and M. Yamaguchi who
contributed to most of our recent production on the subject which was
reported in these talks. A.M. thanks the organizers for the
stimulating settling in which the workshop and the symposium took
place. The work of A.M. was partly supported by the TMR project
``Beyond the Standard Model'' contract number ERBFMRX CT96 0090.  L.S.
acknowledges the support of the German Bundesministerium f{\"u}r
Bildung und Forschung under contract 06 TM 874 and DFG Project Li
519/2-2.

\end{document}